\documentclass[10pt,letterpaper]{article}
\usepackage[top=0.85in,left=2.75in,footskip=0.75in,marginparwidth=2in]{geometry}

\usepackage[utf8]{inputenc}

\usepackage{cite}

\usepackage{nameref,hyperref}

\usepackage[right]{lineno}

\usepackage{microtype}
\DisableLigatures[f]{encoding = *, family = * }

\raggedright
\usepackage{changepage}

\usepackage[aboveskip=1pt,labelfont=bf,labelsep=period,singlelinecheck=off]{caption}

\makeatletter
\renewcommand{\@biblabel}[1]{\quad#1.}
\makeatother

\usepackage{lastpage,fancyhdr,graphicx}
\usepackage{epstopdf}
\fancyheadoffset[L]{2.25in}
\fancyfootoffset[L]{2.25in}

\usepackage{color}

\definecolor{Gray}{gray}{.25}

\usepackage{graphicx}

\usepackage{sidecap}

\usepackage{wrapfig}
\usepackage[pscoord]{eso-pic}
\usepackage[fulladjust]{marginnote}
\reversemarginpar

\begin{document}
\vspace*{0.35in}

\begin{flushleft}
{\Large
\textbf\newline{MicroMegascope}
}
\newline
\\
Luca Canale,
Axel Laborieux,
Agasthya Aroul Mogane,
Laetitia Jubin,
Jean Comtet,
Antoine Lainé,
Lydéric Bocquet,
Alessandro Siria,
Antoine Niguès\textsuperscript{*}.
\\
\bigskip
Laboratoire de Physique Statistique de l'Ecole Normale Superi\'eure, UMR CNRS 8550, PSL Research University, 24 Rue Lhomond 75005 Paris, France
\\
\bigskip
* antoine.nigues@lps.ens.fr

\end{flushleft}

\section*{Abstract}
Atomic Force Microscopy (AFM) allows to reconstruct the topography of surface with a resolution in the nanometer range. The exceptional resolution attainable with the AFM makes this instrument a key tool in nanoscience and technology. The core of the set-up relies on the detection of the mechanical properties of a micro-oscillator when approached to a sample to image. Despite the fact that AFM is nowadays a very common instrument for research and development applications, thanks to the exceptional performances and the relative simplicity to use it, the fabrication of the micrometric scale mechanical oscillator is still a very complicated and expensive task requiring a dedicated platform. Being able to perform atomic force microscopy with a macroscopic oscillator would make the instrument more versatile and accessible for an even larger spectrum of applications and audiences. 
We present for the first time atomic force imaging with a centimetric oscillator. We show how it is possible to perform topographical images with nanometric resolution with a grams tuning fork. The images presented here are obtained with an aluminum tuning fork of centimeter size as sensor on which an accelerometer is glued on one prong to measure the oscillation of the resonator.  In addition to the stunning sensitivity, by imaging both in air and in liquid, we show the high versatility of such oscillator. The set up proposed here can be extended to numerous experiments where the probe needs to be heavy and/or very complex as well as the environment.


\section*{Introduction}
Atomic Force Microscope (AFM) is a powerful instrument to both reconstruct topography at the nanoscale of a sample surface and measure interactions at nanoscale. Since its invention in 1986 by Binning and Rohrer \cite{Binnig1986}, lots of efforts have been dedicated to this instrument to improve its capacities \cite{Giessibl1995,Giessibl2003,Extance2018} and to make it affordable; nowadays AFM is an essential tool for a large spectrum of application ranging from condensed matter and soft matter to biological science \cite{McGraw2017,Comtet2016,Dufrene2017}. In the mostly used configuration, a tiny mechanical oscillator, externally excited at the resonant frequency, is scanned over a surface: the interaction forces between a sharp tip at the apex of the oscillator and the sample induce a change in the mechanical properties of the oscillator itself. Keeping constant the interactions between the tip and the surface during an image allows then to reconstruct the sample topography with a resolution in the nanometer range. While the spatial resolution is only limited by the size of the tip, the ability to detect interaction forces relies totally on the oscillator that is the force probe of AFM. The standard and most common force probe is a cantilever with micro- and sub-micrometer dimensions. The quest to ultimate force sensitivities has pushed the development of alternative kind of probes such as unidimensional objects like nanowires and nanotubes and suspended membranes made of graphen and other 2D materials \cite{Nigues2014a,Gloppe2014,Poncharal1999,Miao2014,DeAlba2016}. While the sensitivity is actually being pushed down to impressive values of zepto-Newton \cite{Chaste2012}, these new probes present important constraints due to the challenges in detection and working conditions and it is not possible to easily move them outside laboratory applications \cite{Nigues2017,verlot2017}. On the other hand, and somehow in contrast to this, it is important to develop force probes that couple high sensitivities together with versatility: in this work we present a new atomic sensor, named MicroMegascope (MiMes), based on a centimetric harmonic oscillator. The advantages of using macroscopic probes is twofold : first, due to its dimensions, it is possible to change the specificity of the probe at convenience: this allows then to study interactions in a variety of geometries ranging from nanometer size tips up to macroscopic spheres or more complex shapes ; Secondly, due to its mass ($\approx$ 100 g) the coupling with macroscopic devices for position measurements doesn't affect the mechanical properties of the tuning fork enough to substantially decrease the force detection performances of the set-up. In addition to a study of the force detection performances of the MiMes and to demonstrate the potentiality of this new sensor, we perform in this work images at the nanoscale of a sample in air and totally immersed in a highly viscous liquid.  

\section*{MiMes: experimental set-up and force sensor properties}
The MiMes is presented in Figure 1a. The core of the microscope is a centimeter-sized tuning fork made of aluminum. The tuning fork has been designed and realized to reproduce the same geometry and dimension ratio between the different elements as in quartz tuning forks widely used in AFM but with a rescaling factor of 20 \cite{Karrai1995}. The prong of the tuning fork is $l=7.5cm$ long, $w=6,8mm$ wide and $t=12mm$ thick. The prong oscillations are detected using an accelerometer directly glued at the extremity of one prong. The oscillation amplitude $A$ is directly proportional to the acceleration $a_{acc}$ measured by the accelerometer such that $A=a_{acc}/f_0^{2}$. The tuning fork coupled to the accelerometer alone represents the MiMes. A similar device has been presented by Bosma et al \cite{Bosma2010}, showing that the topography of a coin surface could be reconstructed with a resolution in the micron range. By increasing the mechanical properties of the force sensor and the displacement detection resolution, we can apply the technique to the field of Atomic Force Microscopy.
\begin{figure*}[!t]
\centering
\captionsetup{justification=centering}
\includegraphics[width=\textwidth]{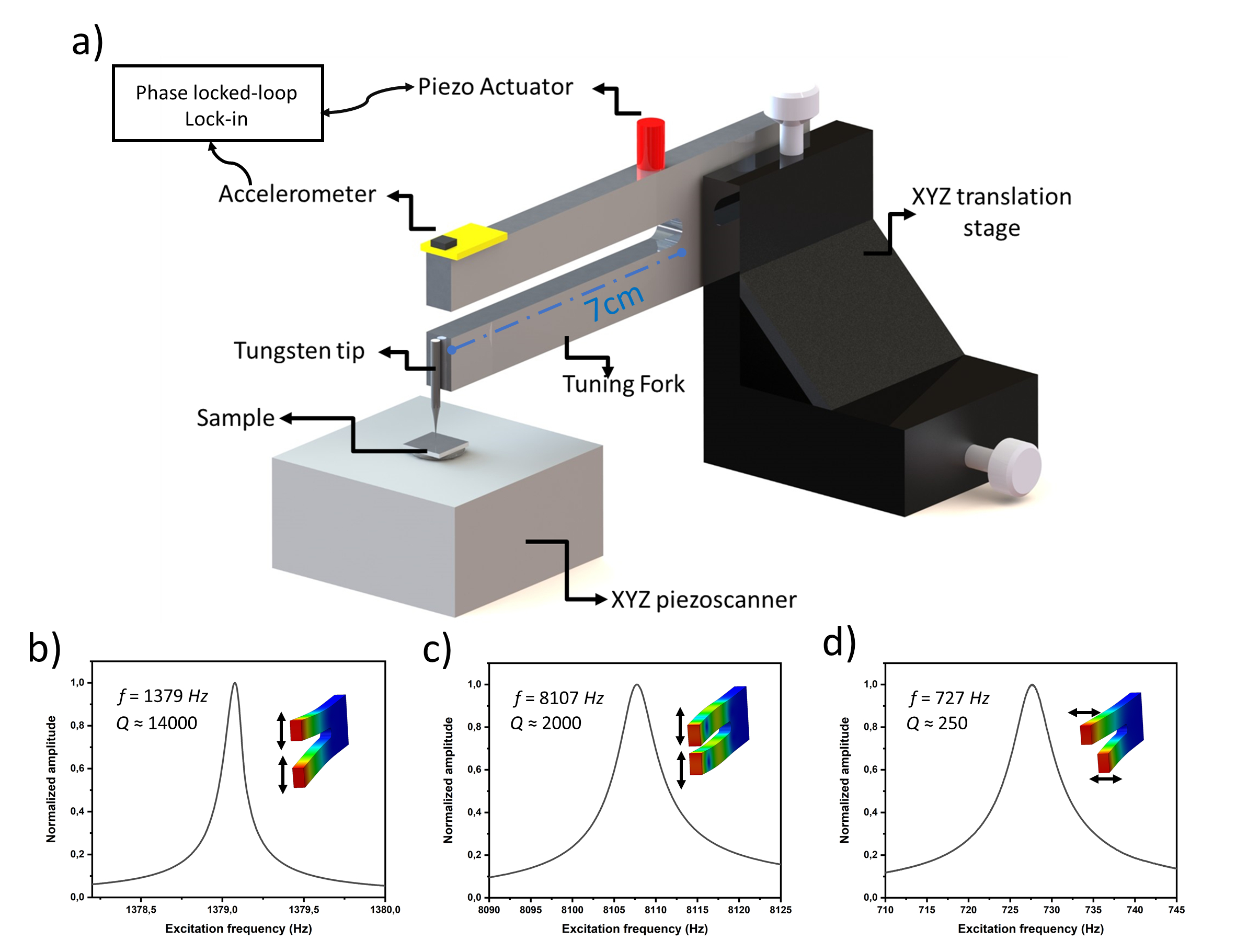}
\caption{Experimental set-up and characterization. a) Schematic drawing of the experimental set-up; a centimetric tuning fork to which a fine tungsten tip is attached is mechanically excited by a piezo at its resonant frequency. These oscillations are measured by an accelerometer glued to the end of one of the tuning fork arms. The signals from the accelerometer are transmitted to a lock-in and a Phase-locked loop to keep both the amplitude and phase between the tuning fork and the excitation constant. The tuning fork is attached to a translation stage for coarse approach and a piezoscanner is used for fine approach and nanoscale imaging of the sample. b),c) and d) Resonance curve for respectively the fundamental and first harmonic of the normal mode and the fundamental harmonic of the tangential mode.}
\end{figure*}\\
The tuning fork and its accelerometer are attached to a XYZ micrometric translation stage for the coarse approach. A piezo-actuator glued to the base of the tuning fork ensures the mechanical excitation. Further, to perform images of sample surfaces, a chemically etched tungsten wire with a radius at the apex of $\approx$ 50 nm is glued at the extremity of one prong . Finally the sample is placed on a three axis piezoscanner with sub-nanometric resolution in displacement (Tritor101 Piezosystemjena). \\
The spring constant $k$ of the tuning fork is given by :
\begin{eqnarray}
k=\frac{Ewt^3}{4l^3}
\end{eqnarray}
where $E$ is the Young modulus for Aluminum, $E=69$ GPa leading to $k=480$ kN/m. The resonance frequency of the fundamental mode is given by :
\begin{eqnarray}
f_0=\frac{\sqrt{k/m_{eff}}}{2\pi}
\end{eqnarray}
where $m_{eff}=0.24\ \rho\times t\times w\times l=3,8$ g is the effective mass of that mode, $\rho=2600$ kg/m$^3$ the mass density of Aluminum. We then obtain $f_0=1788Hz$. A finite element study performed with COMSOL enables to test the mechanical parameters and characteristics of the tuning fork.
\begin{figure*}[!t]
\centering
\captionsetup{justification=centering}
\includegraphics[width=\textwidth]{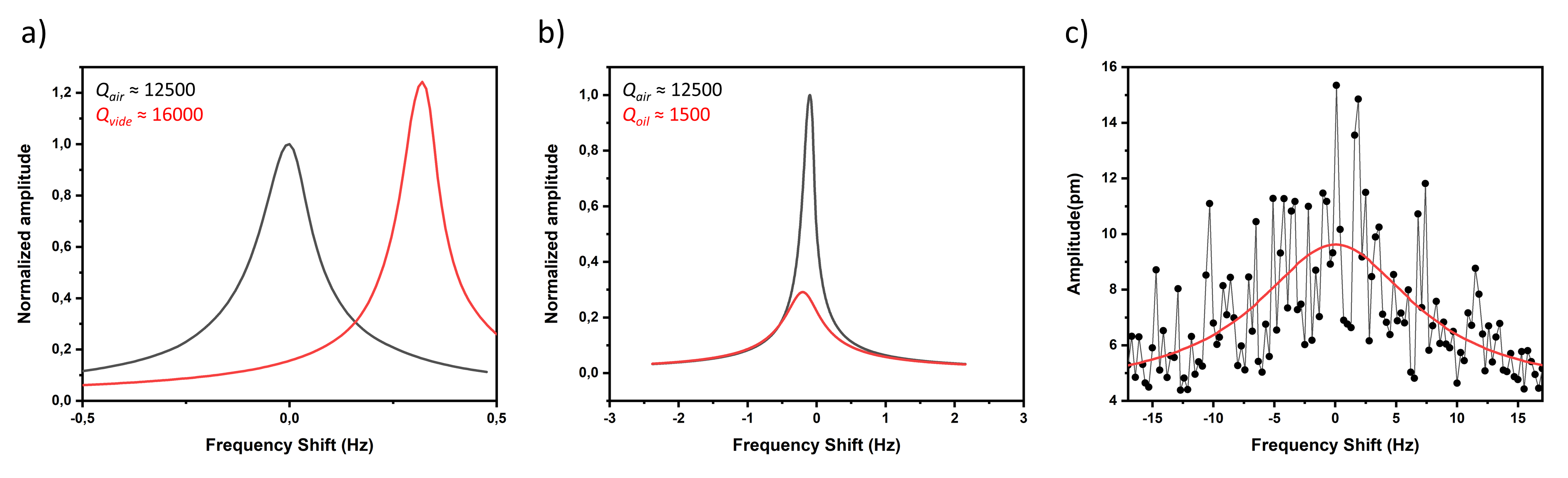}
\caption{Mechanical response of the MiMes. a) Resonance curves in air (black) and in vacuum (red). b) Resonance curves in air (black) and with the tungsten tip immersed in highly viscous liquid (red). c) Thermal noise of the MiMes detected through the accelerometer.}
\end{figure*}\\
In figure 2, we show the mechanical response of the tuning fork around the fundamental resonant frequency. Despite its size, the macroscopic tuning fork is characterized by a low intrinsic dissipation and a large quality factor up to $\approx$ 10000 in air, allowing the detection of the oscillation amplitude of the tuning fork down to its thermal motion and to oscillation amplitudes of the order of 15 pm. It is worth to compare this value to the thermal variance expected for such tuning fork, $\Delta x^2_{th}=k_BT/m_{eff}\omega^{2}_{0}$, with $k_B$ Boltzmann's constant and $T=300K$ the ambient temperature. We find $\Delta x^2_{th}=(10pm)^2$, in very good agreement with the experimental value measured above.\\ 
At this point it is now important to determine the force sensitivity of our tuning fork. The force sensitivity of an oscillator in a certain bandwidth $B$ is given by \cite{rugar1997}:
\begin{eqnarray}
F_{min}=\sqrt{\frac{wt^2}{lQ}}(E\rho)^{(1/4)}(k_BTB)^{(1/2)}
\end{eqnarray}
Inserting the parameters of the MiMes in eq. 3 we obtain a minimal force detection of 21 pN$/\sqrt{Hz}$. This minimum achievable force can be improved by a factor of one to two by improving the intrinsic dissipation of our tuning fork, by changing its manufacturing material and/or by working under vacuum conditions, as shown in figure 2a. However the force sensitivity obtained for the chosen configuration is already compatible with near field force measurement and atomic force microscopy.\\
The operating principle of a tuning fork as a force sensor is as follows:  when excited by an external sinusoidal force $F_{ext}(\omega)=F_{ext}e^{i\omega t}$, the tuning fork behaves in first approximation as a spring-mass system with oscillation amplitude and phase with respect to the excitation given by:
\begin{eqnarray}
A(\omega)=\frac{F_{ext}}{\sqrt{m^2_{eff}(\omega^{2}_{0}-\omega^{2})^2+\gamma^2\omega^{2}}} \\
\phi(\omega)=\arctan\left(\frac{\gamma\omega}{m_{eff}(\omega^{2}_{0}-\omega^{2})}\right)
\end{eqnarray}
with $\gamma$ the damping factor.
As the interaction of the oscillator with its environment is modified, one observes a change in both the frequency and the amplitude at resonance. The shift in resonance frequency $\delta f$ is related to the conservative force response, whereas the broadening of the resonance (change of quality factor $Q_0\rightarrow Q_1$) is related to dissipation:
\begin{eqnarray}
\frac{\partial F}{\partial r}= 2 k \frac{\delta f}{f_{0}} \\
F_{D} = \frac{kA}{\sqrt{3}} \left( \frac{1}{Q_0} -\frac{1}{Q_1} \right)
\end{eqnarray}
During the experiments, measurements and controls are performed in Real-time by a complete Specs-Nanonis package (RT5, SC5 and OC4) and  two feedback loops enable to work at the resonance and maintain constant the oscillation amplitude $A$ by changing the voltage amplitude applied to the piezo actuator.\\ 
From equations 6 and 7, the ability to detect the interaction forces with a sample is determined by the spring constant of the force sensor. Even if the stiffness of the tuning fork is an order of magnitude larger than the classical quartz tuning fork and above the range of optimal stiffness values for frequency modulation microscopy \cite{Giessibl2013,Bosma2010}, we will show in the following that this is not a limiting factor for obtaining nanometrically resolved image of the surface of a sample.
\begin{figure*}[!t]
\centering
\captionsetup{justification=centering}
\includegraphics[width=\textwidth]{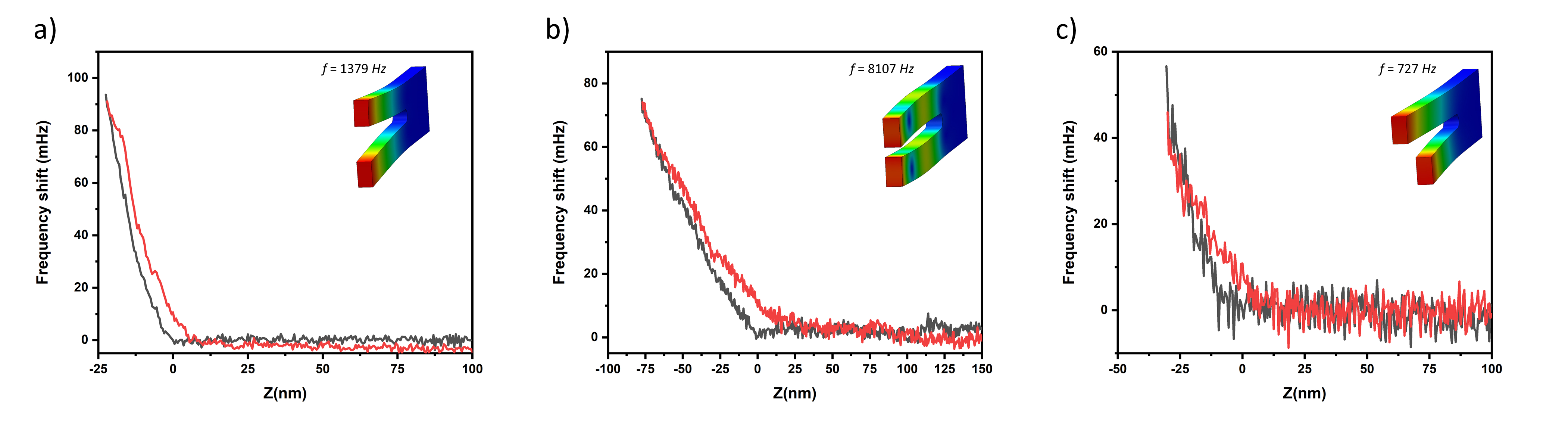}
\caption{Force curves obtained with a tungsten tip on a Silicon dioxide substrate at different frequencies ; a) Fundamental harmonics of normal mode ($\approx 1350Hz$). b) First harmonics of normal mode ($\approx 8100Hz$). c) Fundamental harmonics of the tangential mode ($\approx 700Hz$). Approach in black, retract in red.}
\end{figure*}

\section*{Results}
To demonstrate the potential of MiMes for force microscopy we initially performed a series of approach-retract curves on a Silicon Dioxide flat surface. The interaction between the apex of a sharp tungsten tip glued at the extremity of one prong and the sample surface are detected by measuring the shift in the resonant frequency. In Figure 3, we present the measurement for the fundamental frequency of the normal mode, the first harmonics of the same mode as well as the fundamental frequency of the tangential mode. These measurements prove that the macroscopic force sensor can detect the near field interaction forces demonstrating that the technique can be applied for AFM applications and friction studies.\\
The AFM images in Figure 4 have been performed in the so-called FM-AFM. In this mode the substrate is scanned  with constant frequency shift, i.e. constant force gradient. The amplitude of vibration $A$ of the tuning fork is kept constant at $10nm$. Figure 4a shows a nanometricaly resolved standard calibration grating with a pitch of $5\mu m$ and depth $180nm$. Notwithstanding the effects inherent in the piezoelectricity of open-loop scanners (creep, hysteresis...), this first image obtained with a centimetric oscillator corresponds in every aspect to the criteria expected with a conventional AFM probe. This irrefutably shows the exceptional sensitivity that our centimetric mechanical oscillator coupled with MEMS detection can achieve. In order to push the nail even deeper and prove the great versatility of MiMes, we proceeded to the imaging of the same surface but completely immersed in a highly viscous liquid, silicone oil (10000 cst). Indeed, it is no longer necessary to specify that AFM imaging in liquid media is still a challenge to this day. When fully immersed, the quality factor of conventional levers decreases drastically, laser detection is deteriorated by beam reflection on the liquid surface and multi-peaks resonances appear making it difficult to distinguish the natural frequency of the lever. Similar problems appear with quartz tuning forks. In our case, our sensor does not interact directly with the liquid, so its properties and sensitivity are not deteriorated. The image shown in Figure 4b does not show any more defects than the one obtained in the air and its realization did not require more technical means than in the air. This completes the demonstration of the versatility, sensitivity and ease of use of the MiMes.
\begin{figure*}[!t]
\centering
\captionsetup{justification=centering}
\includegraphics[width=\textwidth]{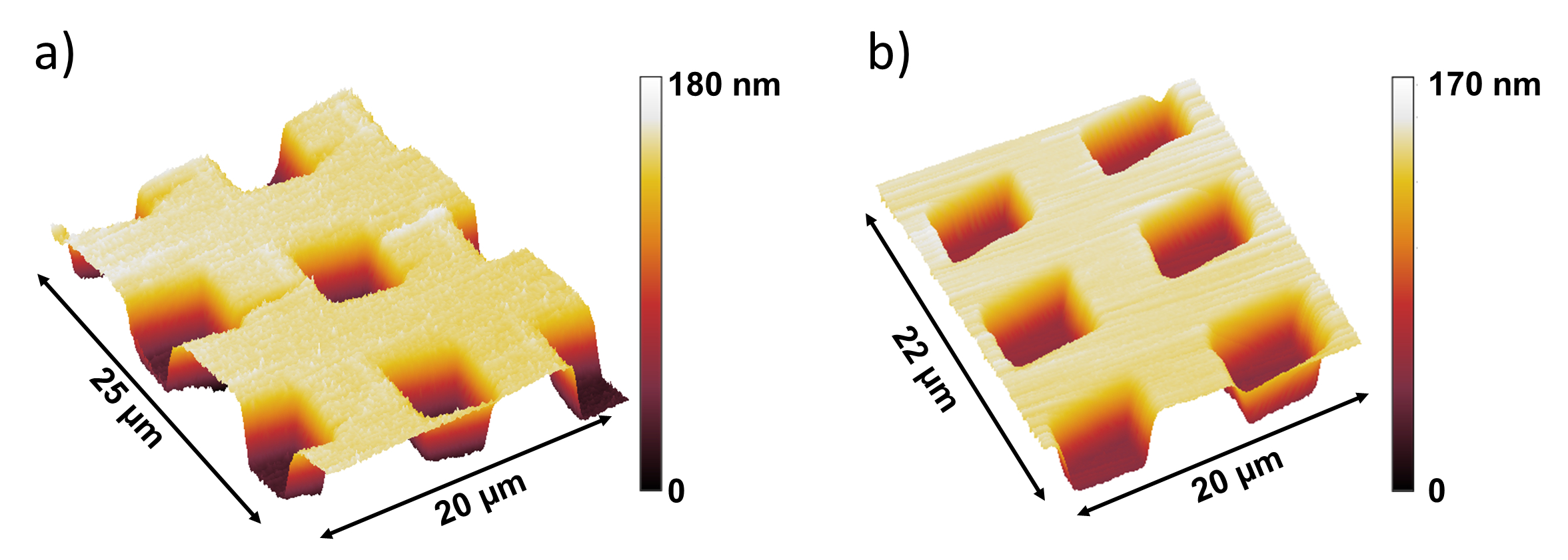}
\caption{Topographical images of calibration grating (pitch $5\mu m$ depth $160n$) a) in air and b) immersed in silicon oil.}
\end{figure*}

\section*{Conclusions and discussions}
In conclusion we have demonstrated that a macroscopic mechanical oscillator can be used as force sensor for atomic force microscopy. Despite the size and mass, the macroscopic tuning fork presents the force sensitivity needed to probe near field surface interactions and image surface topography with nanometer resolution, provided a suitable tip is attached at the extremity of one prong. We have performed atomic force measurements and imaging in air and in high viscous liquid showing no remarkable effect of the environment on the image quality. Because of its size, the force sensor can support a macroscopic tip immersed in the fluid while keeping the force sensor in air, maintaining the mechanical properties and force sensitivity unperturbed.\\
It is worth citing that beyond the performances for Atomic Force Microscopy, centimeter size tuning forks can be implemented as force sensor for a broad spectrum of applications. The possibility to change the probe size and geometry, ranging from nanometric tips to macroscopic spheres, allows to perform measurements of surface interactions in analogy with dynamical Surface Force Apparatus (d-SFA) \cite{Restagno}. Finally the large mechanical stability and low intrinsic dissipation coupled with the possibility to detect orthogonal mechanical resonances, can open the way to the development of a new class of instruments for the measurement of friction phenomena in complex media \cite{Comtet2017}.


\begin{thebibliography}{10}
\bibitem{Binnig1986} Binnig, J., Quate, C.F., Gerber, C., Atomic force microscope, Phys. Rev. Lett., 56, 930-934 (1986) 
\bibitem{Giessibl1995} Giessibl, F. J., Atomic resolution of the silicon (111)-(7x7) surface by atomic force microscopy, Science, 13, 68-70 (1995)
\bibitem{Giessibl2003} Giessibl, F. J., Advances in atomic force microscopy, Rev. Mod. Phys., 3, 949-983 (2003)
\bibitem{Extance2018} Extance, A. (2018). The Atomic-Force Revolution. Nature, 555, 545. https://doi.org/10.1038/d41586-018-03305-2
\bibitem{McGraw2017}McGraw, J., Nigu\`es, A., Chennevière, A., and Siria, A. (2017). Contact dependence and velocity crossover in friction between microscopic solid/solid contacts. Nano Letters, acs.nanolett.7b03076. https://doi.org/10.1021/acs.nanolett.7b03076
\bibitem{Comtet2016} Comtet, J., Nigu\`es, A., Kaiser, V., Bocquet, L., and Siria, A. (2016). Nanoscale capillary freezing of ionic liquids confined between metallic interfaces and the role of electronic screening. Nature Materials, 1(March). https://doi.org/10.1038/nmat4880
\bibitem{Dufrene2017} Dufr\^ene, Y. F., Ando, T., Garcia, R., Alsteens, D., Martinez-Martin, D., Engel, A., Müller, D. J. (2017). Imaging modes of atomic force microscopy for application in molecular and cell biology. Nature Nanotechnology, 12(4), 295–307. https://doi.org/10.1038/nnano.2017.45
\bibitem{Nigues2014a} Nigu\`es, A., Siria, A., Verlot, P. Dynamical Backaction Cooling with Free Electrons, Nature Communications, doi:10.1038/ncomms9104 (2015)
\bibitem{Gloppe2014} Gloppe, A., Verlot, P., Dupont-Ferrier, E., Kuhn, A. G., Siria, A., Poncharal, P., Bachelier, G., Vincent, P., Arcizet, O., Bidimensional nano-optomechanics and topological backaction in a non-conservative radiation force field, Nature Nanotechnology, 9, 920-926 (2014)
\bibitem{Poncharal1999} Poncharal, P., Wang, Z., Ugarte, D., de Heer W.A., Electrostatic deflections and electromechanical resonances of carbon nanotubes, Science, 283, 1513-1516 (1999)
\bibitem{Miao2014} Miao, T., Yeom, S., Wang, P., Standley, B., Bockrath, M., Graphene nanoelectromechanical systems as stochastic-frequency oscillators, Nano Letters, 6, 2982-2987 (2014)
\bibitem{DeAlba2016} De Alba, R., Massel, F., Storch, I. R., Abhilash, T. S., Hui, A., McEuen, P. L., Craighead, H. G., Parpia, J. M., Tunable phonon cavity coupling in graphene membranes, Nature Nanotechnology, 11, 741-746 (2016)


\bibitem{Chaste2012} Chaste, J., Eichler, A., Moser, J., Ceballos, G., Rurali, R., Bachtold, A., A nanomechanical mass sensor with yoctogram resolution, Nature Nanotechnology, 5, 301-304 (2012)



\bibitem{Nigues2017} Siria, A., and Nigu\`es, A. (2017). Electron beam detection of a Nanotube Scanning Force Microscope. Scientific Reports, 7(September), 1–6. https://doi.org/10.1038/s41598-017-11749-1
\bibitem{verlot2017} Tsioutsios, I., Tavernarakis, A., Osmond, J., Verlot, P., and Bachtold, A. (2017). Real-Time Measurement of Nanotube Resonator Fluctuations in an Electron Microscope. Nano Letters, 17(3), 1748–1755. https://doi.org/10.1021/acs.nanolett.6b05065


\bibitem{Karrai1995} Karrai, K., Grober, R. D., Piezoelectric tip-sample distance control for near field optical microscopes, Appl. Phys. Lett., 14, 1842-1844 (1995)

\bibitem{rugar1997}Stowe, T. D., Yasumura, K., Kenny, T. W., Botkin, D., Wago, K., and Rugar, D. (1997). Attonewton force detection using ultrathin silicon cantilevers. Applied Physics Letters, 71(2), 288–290. https://doi.org/10.1063/1.119522


\bibitem{Giessibl2013}Giessibl, F., Pielmeier, F., Eguchi, T., An, T., and Hasegawa, Y. (2011). Comparison of force sensors for atomic force microscopy based on quartz tuning forks and length-extensional resonators. Physical Review B, 84(12), 1–15. https://doi.org/10.1103/PhysRevB.84.125409

\bibitem{Bosma2010} Bosma, E., Offerhaus, H. L., Van Der Veen, J. T., Segerink, F. B., and Van Wessel, I. M. (2010). Large scale scanning probe microscope: Making the shear-force scanning visible. American Journal of Physics, 78(6), 562. https://doi.org/10.1119/1.3319657


\bibitem{Restagno} Restagno, F., Crassous, J., Charlaix, É., Cottin-Bizonne, C., and Monchanin, M. (2002). A new surface forces apparatus for nanorheology. Review of Scientific Instruments, 73(6), 2292. https://doi.org/10.1063/1.1476719

\bibitem{Comtet2017} Comtet, J., Chatté, G., Nigu\`es, A., Bocquet, L., Siria, A., and Colin, A. (2017). Pairwise frictional profile between particles determines discontinuous shear thickening transition in non-colloidal suspensions. Nature Communications, 8. https://doi.org/10.1038/ncomms15633





\end{thebibliography}
\end{document}